\documentclass{aa}
\usepackage{graphicx}
\def\LB{$\lambda$ Bootis }
\begin{document}

\title{A study of $\lambda$ Bootis type stars in the wavelength region beyond 7000\,\AA
\thanks{Based on observations with ISO, an ESA project with instruments funded by 
ESA Member States (especially the PI countries: France, Germany, the Netherlands and 
the United Kingdom) and with the participation of ISAS and NASA; and observations at 
the Heinrich-Hertz-Telescope (HHT, operated by the the Submillimeter Telescope Observatory).
}}
\author{E.~Paunzen\inst{1,2}, I.~Kamp\inst{3}, W.W.~Weiss\inst{1}, 
H.~Wiesemeyer\inst{4}}

\offprints{E.~Paunzen}

\institute{Institut f\"ur Astronomie der Universit\"at Wien,
           T\"urkenschanzstr. 17, A-1180 Wien, Austria \\
		   e-mail: ernst.paunzen@univie.ac.at, weiss@astro.univie.ac.at
\and       Zentraler Informatikdienst der Universit\"at Wien,
           Universit\"atsstr. 7, A-1010 Wien, Austria
\and	   Leiden Observatory,
		   Niels Bohrweg 2, PO Box 9513, 2330 RA Leiden, The Netherlands \\
		   e-mail: kamp@strw.leidenuniv.nl
\and	   IRAM Grenoble, 300 Rue de la Piscine, Domaine Universitaire, 
		   F-38406 St. Martin d'H{\`e}res Cedex, France \\
		   e-mail: wiesemey@iram.fr}

\date{Received 2002; accepted 2003}
\titlerunning{$\lambda$ Bootis type stars beyond 7000\,\AA}{}
\abstract{The group of \LB type
stars comprises late B- to early F-type, Population\,I objects which
are basically metal weak, in particular the Fe group elements, but
with the clear exception of C, N, O and S. One of the theories to explain the
abundance pattern of these stars involves circumstellar or interstellar matter
around the objects. Hence, we have compiled all available data
from the literature of well established members of the \LB group redward of 7000\,\AA\,
in order to find evidence for matter around these objects. Furthermore,
we present unpublished ISO as well as submillimeter continuum and CO (2\,$-$\,1) line
measurements to complete the data set.
In total, measurements for 34 (26 with data redward of 20\,$\mu$m) well established \LB stars 
are available.
There is evidence for an infrared excesses in six stars (HD~31295, HD~74873, HD~110411, 
HD~125162,
HD~198160/1 and HD~210111) and two are doubtful cases (HD~11413 and HD~192640)
resulting in a percentage of 23\,\% (excluding the two doubtful cases). 
Dust models for these objects show fractional dust luminosities comparable 
to the Vega-type stars and slightly higher dust temperatures.
ISO-SWS spectroscopy for HD~125162 and HD~192640 resulted in the detection
of pure stellar H\,{\sc i} lines ruling out an active accretion disk (as found
for several Herbig Ae/Be stars) around these objects. The
submillimeter measurements gave only upper limits for the line and continuum
fluxes.
}

\maketitle

\section{Introduction}

The group of \LB stars comprise of true
Population\,I, late B to early F-type stars, 
with moderate to extreme (up to a factor 100) surface underabundances of most
Fe-peak elements and solar abundances of lighter elements (C, N, O, and S).
Only a maximum of about 2\% of all objects in the
relevant spectral domain are believed to be \LB type stars (Paunzen 2001, Gray \& Corbally
2002).

One of the most promising theories explaining the \LB phenomenon is the diffusion/accretion
theory (Turcotte \& Charbonneau 1993, Andrievsky \& Paunzen 2000, Kamp \& Paunzen 2002,
Turcotte 2002).
Within this model, interstellar or circumstellar gas is accreted by the star.
Hence, a circumstellar disk or shell should exist around \LB stars which
manifests itself in an infrared (IR hereafter) excess. 

The first near-infrared data for well established \LB stars was 
already published by Oke (1967). He lists spectrophotometry for
HD~31295, HD~125162 and HD~192640 up to 10870\,\AA. 
Data for wavelengths beyond 10\,$\mu$m became
available with the launch of the IRAS-satellite. The IR excesses
detected for Vega, $\epsilon$\,Eridani,
$\alpha$\,Piscis Austrini and $\beta$\,Pictoris (Aumann et al. 1984)
have been interpreted as due to cold circumstellar disks or shells around
these stars. Later, the disk of $\beta$\,Pictoris was directly imaged
in the IR revealing a solar system-sized dust disk extending to at least
1100\,AU (Smith \& Terrile 1984). A simple model of this system indicates
a central star (spectral type: A5V), encircled by an ionized gaseous inner
disk and an outer dust disk. Aumann (1985) established
the group of Vega-like stars containing main-sequence (MS hereafter) stars also showing an
infrared excess (defined by the IRAS colors at
12 and 60\,$\mu$m: [12]\,$-$\,[60]\,$>$\,1\,mag) similar to the prototype Vega.
Many Vega-like stars are in fact much cooler than Vega (A0\,V) itself.

Several papers (Gerbaldi 1991, King 1994, Andrillat et al. 1995) were devoted
to the group of \LB stars presenting data in the IR wavelength region but none
of them presented a comprehensive
analysis of all available data. Furthermore, the ISO satellite brought a flood of
new data for group members. Fajardo-Acosta et al. (1999) presented data for three
\LB stars, namely HD~11413, HD~110411 and HD~192640.

In Sect.\,\ref{lit_all}, we summarize the current literature status on the \LB
phenomenon focusing especially on circumstellar dust and gas. We have 
reanalyzed in Sect.\,\ref{strat} all available measurements in the wavelength range 
beyond 7000\,\AA, using the newest available calibrations 
(especially important for ISO measurements) and present data for 36 well established 
\LB stars. This
starting wavelength was chosen such that a good estimate for fitting a blackbody 
radiation curve could be made. The final goal of this work is to derive the incidence of objects
with IR excess and to compare the observed fluxes for
\LB type objects with those of models to derive first estimates for the dust
temperature and the optical depth of the circumstellar and/or interstellar
material (Sect.\,\ref{sect_res}). 

\begin{table}
\caption[]{Presence of infrared excess and/or narrow absorption lines 
           towards the well established, 
           apparent single type \LB stars. A question mark indicates doubtful
           cases. The projected rotational velocities were taken from Paunzen et al. (2002).
		   We have indicated the origin of gas features either as interstellar (IS) or
		   circumstellar (CS) from the literature as listed in Sect.\,\ref{obs_gas}.} 
\label{lb_cs}
\begin{tabular}{llrllr}
\hline
\multicolumn{1}{c}{HD}	 &   \multicolumn{1}{c}{HR}  &     \multicolumn{1}{c}{IRAS}    & 
dust & gas & \multicolumn{1}{c}{$v$\,sin\,$i$} \\
\hline
11413    &  541  & F01489$-$5027 &  y?  & CS? & 125 \\
31295    & 1570  &  04521+1004 &  y  & & 115 \\
74873    & 3481  & F08441+1217 &  y  & & 130 \\
110411   & 4828  & 12394+1030  &  y  & & 165 \\
125162   & 5351  & 14144+4619  &  y  &  CS? & 115 \\
142994   &       &  &     &  CS & 180 \\
183324   & 7400  &  &     &  CS? & 90 \\
192640   & 7736  & 20126+3639  &  y? &  IS & 80 \\
193256   & 7764C &  &     &  IS & 250 \\
193281   & 7764A & F20173$-$2921 &     &  IS & 95 \\
198160   & 7959  & F20474$-$6237 &  y  & CS? & 200 \\
210111   & 8437  & F22058$-$3322 &  y  & & 55 \\
221756   & 8947  & 23321+3957  &  y  &  CS & 105 \\
\hline
\end{tabular}
\end{table}

\section{The \LB phenomenon and observations in the IR-region} \label{lit_all}

In principle, three models were developed to explain the \LB
phenomenon: 1) the diffusion/mass-loss; 2) the diffusion/accretion
and 3) the binarity model. Since the second model is the one
relevant for this paper, we describe the other two just briefly.

The diffusion/mass-loss theory was formulated by Michaud \& Charland (1986). They
modified the highly successful diffusion model 
for the AmFm phenomenon in order to account for stellar mass-loss. 
AmFm stars are Population I, nonmagnetic
MS stars with underabundances of Ca and Sc, but large overabundances
of most heavier elements (up to a factor of 500). This abundance pattern is
explained by gravitional settling of He
which leads to the disappearance of the outer convection zone associated with 
He-ionization. Later, Charbonneau (1993) showed that (at any time 
of the MS evolution) even moderate equatorial rotational velocities 
of 50\,km\,s$^{-1}$ seem to prevent  the appearance of the underabundance pattern,
because meridional circulation effectively mixes the stellar atmosphere. Beside the
estimation of stellar ages for members of the \LB group, so far no other observational
tests are suggested for this model.

The binarity theory explains the apparent spectral peculiarities either 
by undetected spectroscopic binary systems (Faraggiana \& Bonifacio 1999) or
by a single type object which is a product of a merging process similar to
blue stragglers (Andrievsky 1997). Both approaches can only account for
the nature of a small number of objects. Faraggiana et al. (2001)
and Marchetti et al. (2001) failed to detect any new spectroscopic binary system
among their observed sample. Iliev et al. (2002) reported about two 
spectroscopic binary systems, which comprise only \LB type objects.

The diffusion/accretion model is so far the most promising theory in order
to explain the \LB phenomenon.
Venn \& Lambert (1990) were the first who noticed the similarity 
between the abundance pattern of \LB stars
and the depletion pattern of the interstellar medium (ISM) and
suggested the accretion of interstellar or circumstellar gas to 
explain the \LB stars. In the ISM metals are underabundant because of their 
incorporation in dust grains or ice mantles around the dust grains. 

Following this, Waters et al. (1992) worked out a scenario, where
the \LB star accretes metal-depleted gas from a
surrounding disk. In this model, the dust grains are blown away 
by radiation pressure and coupling between dust and gas is
negligible. Considering the spectral-type of \LB stars,
the gas in the disk remains neutral and hence does not experience
significant direct radiation pressure. 

\begin{table*}[htb]
\caption{Summary of fluxes for \LB type stars red-ward of 7000\,\AA.
The data are given in Jansky (Jy), 1\,Jy = 10$^{26}$\,W\,m$^{-2}$\,Hz$^{-1}$
= 10$^{23}$\,erg\,s$^{-1}$\,cm$^{-2}$\,Hz$^{-1}$ = 
$\lambda^{2}\cdot$3.336$\cdot$10$^{4}$\,erg\,s$^{-1}$\,cm$^{-2}$\,\AA$^{-1}$. 
In parenthesis are the errors in the final digits of the corresponding quantity;
{\it italized} values are only {\it upper limits}. Photometric systems: Johnson (J), 
spectrophotometry from Oke (1967),
13\,color system (13c), IRAS and ISO. Note for the ISO measurements: there are chopped (C) 
and staring mode (S) observations. All star numbers are according to the HD catalogue.}
\label{result}
\begin{scriptsize}
\begin{tabular}{llcccccccccc}
\hline
System & \multicolumn{1}{c}{$\lambda_{c}$ [$\mu$m]} & 319 & 7908 &
11413 & 15165 & 23392 & 24472 & 30422 & 31295 & 35242 \\
\hline
J\,(R) & 0.7 & -- & -- & -- & -- & -- & -- & -- & 44.3(2.2) & -- \\
13c & 0.7241 & -- & -- & -- & -- & -- & -- & -- & 39.0(2.0) & -- \\
13c & 0.8 & -- & -- & -- & -- & -- & -- & -- & 35.1(1.8) & -- \\
13c & 0.8584 & -- & -- & -- & -- & -- & -- & -- & 33.6(1.7) & -- \\
J\,(I) & 0.9 & -- & -- & -- & -- & -- & -- & -- & 35.7(1.8) & -- \\
13c & 0.9831 & -- & -- & -- & -- & -- & -- & -- & 33.3(1.7) & -- \\
Oke & 0.995 & -- & -- & -- & -- & -- & -- & -- & -- & -- \\
Oke & 1.0256 & -- & -- & -- & -- & -- & -- & -- & 34.9(1.7) & -- \\
Oke & 1.04 & -- & -- & -- & -- & -- & -- & -- & -- & -- \\
Oke & 1.061 & -- & -- & -- & -- & -- & -- & -- & 33.9(1.7) & -- \\
Oke & 1.0796 & -- & -- & -- & -- & -- & -- & -- & 33.9(1.7) & -- \\
Oke & 1.087 & -- & -- & -- & -- & -- & -- & -- & 33.0(1.7) & -- \\
13c & 1.1084 & -- & -- & -- &-- & -- & -- & -- & 30.0(1.5) & -- \\
J\,(J) & 1.25 & -- & 3.37(1) & -- & -- & 0.84(1) & 4.42(2) & 7.68(4) & -- & -- \\
J\,(H) & 1.65 & 6.79(3) & 2.53(2) & -- & -- & 0.56(1) & 3.56(2) & 5.56(4) & -- & -- \\
J\,(K) & 2.2 & 4.60(2) & 1.77(1) & -- & -- & 0.38(1) & 2.32(1) & 3.68(1) & -- & -- \\
ISO P1(C) & 3.29 & -- & -- & -- & -- & -- & -- & -- & -- & -- \\
ISO P1(C) & 3.6 & -- & -- & -- & -- & -- & -- & -- & -- & -- \\
ISO P1(S) & 3.6 & 1.38(1) & -- & 2.24(5) & -- & -- & -- & -- & -- & -- \\
J\,(L) & 3.8 & -- & -- & -- & -- & -- & -- & -- & -- & -- \\
J\,(M) & 4.7 & -- & -- & -- & -- & -- & -- & -- & -- & -- \\
ISO P1(C) & 4.85 & -- & -- & -- & -- & -- & -- & -- & -- & -- \\
ISO P1(C) & 11.5 & -- & -- & -- & -- & -- & -- & -- & -- & -- \\
ISO P1(S) & 11.5 & 0.21(2) & -- & 0.36(3) & -- & -- & -- & -- & -- & -- \\
IRAS & 12 & -- & -- & 0.20(4) & 0.13(3) & -- & 0.10(2) & 0.16(3) & 0.50(6) & 0.11(2) \\
ISO P2(C) & 20 & -- & -- & -- & -- & -- & -- & -- & -- & -- \\
ISO P2(S) & 20 & 0.09(13) & -- & -- & -- & -- & -- & -- & -- & -- \\
IRAS & 25 & -- & -- & {\it 0.09} & {\it 0.13} & -- & {\it 0.06} & {\it 0.06} & 0.19(4) & {\it 0.11} \\
ISO P2(C) & 25 & -- & -- & -- & -- & -- & -- & -- & -- & -- \\
IRAS & 60 & -- & -- & {\it 0.13} & {\it 0.07} & -- & {\it 0.05} & {\it 0.05} & 0.25(5) & {\it 0.17} \\
ISO P3(S) & 60 & -- & -- & -- & -- & -- & -- & -- & -- & -- \\
ISO C1(C) & 60 & {\it 0.05} & -- & 0.52(28) & -- & -- & -- & -- & -- & -- \\
IRAS & 100 & -- & -- & {\it 0.57} & {\it 0.20} & -- & {\it 0.22} & {\it 0.17} & {\it 0.92} & {\it 0.70} \\
ISO P3(C) & 100 & -- & -- & -- & -- & -- & -- & -- & -- & -- \\
\hline
System & \multicolumn{1}{c}{$\lambda_{c}$ [$\mu$m]} & 74873 &
84123 & 90821 & 91130 & 101108 & 106223 & 110377 & 110411 & 111604 \\
\hline
J\,(R) & 0.7 & -- & -- & -- & -- & -- & -- & -- & -- & -- \\
13c & 0.7241 & -- & -- & -- & -- & -- & -- & -- & 32.2(1.6) & -- \\
13c & 0.8 & -- & -- & -- & -- & -- & -- & -- & 29.4(1.5) & -- \\
13c & 0.8584 & -- & -- & -- & -- & -- & -- & -- & 27.5(1.4) & -- \\
J\,(I) & 0.9 & -- & -- & -- & -- & -- & -- & -- & -- & -- \\
13c & 0.9831 & -- & -- & -- & -- & -- & -- & -- & 27.6(1.4) & -- \\
Oke & 0.995 & -- & -- & -- & -- & -- & -- & -- & -- & -- \\
Oke & 1.0256 & -- & -- & -- & -- & -- & -- & -- & -- & -- \\
Oke & 1.04 & -- & -- & -- & -- & -- & -- & -- & -- & -- \\
Oke & 1.061 & -- & -- & -- & -- & -- & -- & -- & -- & -- \\
Oke & 1.0796 & -- & -- & -- & -- & -- & -- & -- & -- & -- \\
Oke & 1.087 & -- & -- & -- & -- & -- & -- & -- & -- & -- \\
13c & 1.1084 & -- & -- &-- & -- & -- & -- & -- & 25.1(1.3) & -- \\
J\,(J) & 1.25 & -- & 5.50(3) & 0.32(3) & 9.14(5) & 0.64(1) & 3.44(1) & -- & 19.5(1.0) & -- \\
J\,(H) & 1.65 & 5.98(5) & 4.38(3) & 0.23(1) & 6.46(3) & 0.45(1) & 2.71(1) & -- & -- & -- \\
J\,(K) & 2.2 & 4.29(3) & 3.10(2) & 0.15(1) & 4.46(3) & 0.30(1) & 1.85(1) & -- & 6.9(3) & 5.50(2) \\
ISO P1(C) & 3.29 & -- & -- & -- & -- & -- & -- & -- & -- & -- \\
ISO P1(C) & 3.6 & -- & -- & -- & -- & -- & -- & -- & -- & -- \\
ISO P1(S) & 3.6 & -- & -- & -- & -- & -- & -- & -- & 3.37(15) & -- \\
J\,(L) & 3.8 & -- & -- & -- & -- & -- & -- & -- & -- & -- \\
J\,(M) & 4.7 & -- & -- & -- & -- & -- & -- & -- & -- & -- \\
ISO P1(C) & 4.85 & -- & -- & -- & -- & -- & -- & -- & -- & -- \\
ISO P1(C) & 11.5 & -- & -- & -- & -- & -- & -- & -- & -- & -- \\
ISO P1(S) & 11.5 & -- & -- & -- & -- & -- & -- & -- & 0.34(10) & -- \\
IRAS & 12 & 0.17(3) & 0.12(2) & -- & 0.20(4) & -- & -- & 0.15(3) & 0.40(8) & 0.24(5) \\
ISO P2(C) & 20 & -- & -- & -- & -- & -- & -- & -- & -- & -- \\
ISO P2(S) & 20 & -- & -- & -- & -- & -- & -- & -- & 0.41(13) & -- \\
IRAS & 25 & 0.12(6) & {\it 0.09} & -- & {\it 0.15} & -- & -- & {\it 0.07} & 0.17(3) & {\it 0.13} \\
ISO P2(C) & 25 & -- & -- & -- & -- & -- & -- & -- & -- & -- \\
IRAS & 60 & {\it 0.10} & {\it 0.05} & -- & {\it 0.09} & -- & -- & {\it 0.07} & 0.15(3) & {\it 0.09} \\
ISO P3(S) & 60 & -- & -- & -- & -- & -- & -- & -- & -- & -- \\
ISO C1(C) & 60 & -- & -- & -- & -- & -- & -- & -- & 0.38(19) & -- \\
IRAS & 100 & {\it 0.47} & {\it 0.31} & -- & {\it 0.31} & -- & -- & {\it 0.39} & {\it 0.43} & {\it 0.25} \\
ISO P3(C) & 100 & -- & -- & -- & -- & -- & -- & -- & -- & -- \\
\hline
\end{tabular}
\end{scriptsize}
\end{table*}
\addtocounter{table}{-1}
\begin{table*}
\begin{scriptsize}
\caption[]{continued.}
\begin{tabular}{llccccccccc}
\\
\hline
System & \multicolumn{1}{c}{$\lambda_{c}$ [$\mu$m]} &
125162 & 125889 & 142703 & 142994 & 149130 & 156954 & 168740 & 170680 &
183324 \\
\hline
J\,(R) & 0.7 & 62.2(3.1) & -- & -- & -- & -- & -- & -- & -- & -- \\
13c & 0.7241 & 65.1(3.3) & -- & -- & -- & -- & -- & -- & -- & -- \\
13c & 0.8 & 58.2(2.9) & -- & -- & -- & -- & -- & -- & -- & -- \\
13c & 0.8584 & 55.5(2.8) & -- & -- & -- & -- & -- & -- & -- & -- \\
J\,(I) & 0.9 & 50.1(2.5) & -- & -- & -- & -- & -- & -- & -- & -- \\
13c & 0.9831 & 55.0(2.8) & -- & -- & -- & -- & -- & -- & -- & -- \\
Oke & 0.995 & 53.3(2.7) & -- & -- & -- & -- & -- & -- & -- & -- \\
Oke & 1.0256 & 51.3(2.6) & -- & -- & -- & -- & -- & -- & -- & -- \\
Oke & 1.04 & 51.8(2.6) & -- & -- & -- & -- & -- & -- & -- & -- \\
Oke & 1.061 & -- & -- & -- & -- & -- & -- & -- & -- & -- \\
Oke & 1.0796 & 49.5(2.5) & -- & -- & -- & -- & -- & -- & -- & -- \\
Oke & 1.087 & -- & -- & -- & -- & -- & -- & -- & -- & -- \\
13c & 1.1084 & 48.5(2.4) & -- & -- & -- & -- & -- & -- & -- & -- \\
J\,(J) & 1.25 & 42.1(7) & 0.33(1) & -- & -- & -- & 2.41(1) & -- & -- & -- \\
J\,(H) & 1.65 & 28.1(6) & 0.26(1) & -- & -- & -- & 1.84(1) & -- & -- & -- \\
J\,(K) & 2.2 & 17.9(8) & 0.18(1) & 5.24(3) & -- & -- & 1.27(1) & -- & -- & -- \\
ISO P1(C) & 3.29 & -- & -- & -- & -- & -- & -- & -- & -- & -- \\
ISO P1(C) & 3.6 & -- & -- & 3.07(88) & -- & -- & -- & -- & -- & -- \\
ISO P1(S) & 3.6 & -- & -- & -- & {\it 0.87} & -- & -- & -- & -- & -- \\
J\,(L) & 3.8 & -- & -- & -- & -- & -- & -- & -- & -- & -- \\
J\,(M) & 4.7 & -- & -- & -- & -- & -- & -- & -- & -- & -- \\
ISO P1(C) & 4.85 & 4.30(34) & -- & 1.54(53) & -- & -- & -- & -- & -- & -- \\
ISO P1(C) & 11.5 & 1.82(23) & -- & 0.56(26) & -- & -- & -- & -- & -- & -- \\
ISO P1(S) & 11.5 & -- & -- & -- & {\it 2.72} & -- & -- & -- & -- & -- \\
IRAS & 12 & 0.79(8) & -- & 0.23(4) & -- & 0.25(5) & -- & 0.18(3) & 0.25(5) & -- \\
ISO P2(C) & 20 & -- & -- & 0.22(13) & -- & -- & -- & -- & -- & 0.17(11) \\
ISO P2(S) & 20 & {\it 0.62} & -- & -- & {\it 2.45} & -- & -- & -- & -- & -- \\
IRAS & 25 & 0.26(4) & -- & {\it 0.09} & -- & 0.07(2) & -- & {\it 0.06} & {\it 0.31} & -- \\
ISO P2(C) & 25 & -- & -- & -- & -- & -- & -- & -- & -- & -- \\
IRAS & 60 & 0.25(5) & -- & {\it 0.02} & -- & {\it 0.01} & -- & {\it 0.08} & {\it 0.08} & -- \\
ISO P3(S) & 60 & -- & -- & -- & {\it 1.02} & -- & -- & -- & -- & -- \\
ISO C1(C) & 60 & -- & -- & -- & -- & -- & -- & -- & -- & 0.02(6) \\
IRAS & 100 & {\it 0.59} & -- & {\it 0.33} & -- & {\it 2.25} & -- & {\it 0.66} & {\it 1.95} & -- \\
ISO P3(C) & 100 & 0.22(7) & -- & -- & -- & -- & -- & -- & -- & -- \\
\hline
System & \multicolumn{1}{c}{$\lambda_{c}$ [$\mu$m]} & 
192640 & 193256 & 193281 & 198160 & 204041 & 210111 & 221756 \\
\hline
J\,(R) & 0.7 & 36.1(1.8) & -- & -- & -- & -- & -- & 16.6(8) \\
13c & 0.7241 & 33.2(1.7) & -- & -- & -- & -- & -- & -- \\
13c & 0.8 & 31.2(1.6) & -- & -- & -- & -- & -- & -- \\
13c & 0.8584 & 29.3(1.5) & -- & -- & -- & -- & -- & -- \\
J\,(I) & 0.9 & 30.2(1.5) & -- & -- & -- & -- & -- & 15.4(8) \\
13c & 0.9831 & 29.2(1.5) & -- & -- & -- & -- & -- & -- \\
Oke & 0.995 & 32.4(1.6) & -- & -- & -- & -- & -- & -- \\
Oke & 1.0256 & 32.7(1.6) & -- & -- & -- & -- & -- & -- \\
Oke & 1.04 & 31.5(1.6) & -- & -- & -- & -- & -- & -- \\
Oke & 1.061 & 33.0(1.7) & -- & -- & -- & -- & -- & -- \\
Oke & 1.0796 & 31.5(1.6) & -- & -- & -- & -- & -- & -- \\
Oke & 1.087 & 33.6(1.7) & -- & -- & -- & -- & -- & -- \\
13c & 1.1084 & 26.5(1.3) & -- & -- & -- & -- & -- & -- \\
J\,(J) & 1.25 & 23.1(1.2) & 2.04(1) & 5.61(4) & -- & -- & -- & -- \\
J\,(H) & 1.65 & 16.8(8) & 1.51(1) & 4.80(6) & -- & -- & -- & -- \\
J\,(K) & 2.2 & 10.4(5) & 1.05(1) & 3.26(5) & -- & -- & -- & -- \\
ISO P1(C) & 3.29 & -- & -- & -- & -- & -- & -- & 2.11(63) \\
ISO P1(C) & 3.6 & -- & -- & -- & -- & 1.23(55) & -- & -- \\
ISO P1(S) & 3.6 & 3.86(15) & 1.38(3) & -- & 3.54(11) & -- & 1.20(2) & 2.36(2) \\
J\,(L) & 3.8 & 3.97(20) & -- & -- & -- & -- & -- & -- \\
J\,(M) & 4.7 & 2.44(12) & -- & -- & -- & -- & -- & -- \\
ISO P1(C) & 4.85 & 1.79(71) & -- & -- & -- & 1.17(50) & -- & 0.91(43) \\
ISO P1(C) & 11.5 & -- & -- & -- & -- & 0.26(19) & -- & -- \\
ISO P1(S) & 11.5 & 0.37(3) & {\it 0.03} & -- & 0.39(6) & -- & 0.17(3) & 0.69(6) \\
IRAS & 12 & 0.44(8) & 0.23(4) & -- & 0.25(5) & 0.10(2) & 0.17(3) & 0.24(5) \\
ISO P2(C) & 20 & -- & -- & -- & -- & 0.07(24) & -- & -- \\
ISO P2(S) & 20 & 0.19(13) & 0.31(13) & -- & 0.27(12) & -- & 0.25(14) & 0.08(13) \\
IRAS & 25 & {\it 0.18} & {\it 0.11} & -- & {\it 0.18} & {\it 0.09} & {\it 0.06} & 0.06(3) \\
ISO P2(C) & 25 & 0.20(28) & -- & -- & -- & 0.02(3) & -- & 0.08(5) \\
IRAS & 60 & {\it 1.82} & {\it 0.04} & -- & {\it 0.20} & {\it 0.04} & {\it 0.08} & {\it 0.11} \\
ISO P3(S) & 60 & -- & -- & -- & -- & -- & -- & -- \\
ISO C1(C) & 60 & -- & -- & -- & 0.31(13) & -- & 0.10(7) & 0.17(8) \\
IRAS & 100 & {\it 43.58} & {\it 0.80} & -- & {\it 0.67} & {\it 0.26} & {\it 0.17} & {\it 0.62} \\
ISO P3(C) & 100 & -- & -- & -- & -- & -- & -- & 0.01(51) \\
\hline
\end{tabular}
\end{scriptsize}
\end{table*}

%\clearpage

Turcotte \& Charbonneau (1993) calculated the abundance evolution in the
outer layers of a \LB star assuming accretion rates
between $10^{-15}$ and $10^{-12}$~M$_\odot$~yr$^{-1}$. Solving
the diffusion equation modified this time by an additional accretion term, they
obtained the time evolution of the Ca and Ti abundance stratification 
with and without stellar rotation. From their calculations, a lower limit 
of $10^{-14}$~M$_\odot$~yr$^{-1}$ is derived for the diffusion/accretion
model to produce a typical \LB abundance pattern. 
\clearpage
Moreover their 
rotating models provide evidence that meridional circulation cannot destroy 
the established accretion pattern for rotational velocities below 
125\,km\,s$^{-1}$. Since diffusion wipes out any accretion pattern within 
$10^6$~yr at the end of accretion large number of \LB stars should show observational
evidence for the presence of circumstellar material.

The scenarios discussed above imply a constraint on the 
evolutionary status of the star, because the existence of a disk or 
a shell has to be explained in the context of stellar evolution. 
Circumstellar disks are thought to exist during the pre-main-sequence
phase of stellar evolution, while a shell can either occur in a 
very early phase of pre-main-sequence evolution or after a stellar 
merger.

Kamp \& Paunzen (2002) proposed only recently a slightly different accretion
scenario for the \LB stars, namely the accretion
from a diffuse interstellar cloud. This scenario works at any 
stage of stellar evolution as soon as the star passes a diffuse 
interstellar cloud.
The interstellar dust grains are blown away by the stellar
radiation pressure, while the depleted interstellar gas is
accreted onto the star. Typical gas accretion rates are between 
$10^{-14}$ and $10^{-10}$~M$_\odot$~yr$^{-1}$ depending on the 
density of the diffuse cloud and the relative velocity between 
star and cloud. 

Any accretion scenario for \LB stars involves
dust grains as a mean to deplete certain metals in the gas phase.
Moreover due to the short timescales of mixing, gas accretion is 
supposed to be ongoing in most cases. Hence, in the framework of the 
diffusion/accretion model a search for the infrared emission from this 
circumstellar or interstellar dust as well as for narrow gaseous circumstellar 
or interstellar lines seems promising. Furthermore the detection of such 
characteristics might lead to the confirmation of the diffusion/accretion model and
put further constraints on its details. 

\subsection{Results of infrared measurements from the literature} \label{res_lit}

The definition of the infrared region is already
quite problematic. We follow
the classical definition of the near-IR (7000\,\AA\, to
1\,$\mu$m) and IR (1\,$\mu$m to 100\,$\mu$m) as given by
Jaschek \& Jaschek (1987). We have divided the already published
results in signs of dust or gas around bona-fide \LB stars.
We have not included results for objects which were probably
misclassified as \LB stars (see Paunzen 2001). In the following, two objects
are discussed, whose nature is rather contrary: HD~38545 and
HD~111786. Both objects were classified as \LB type by Gray (1988). HD~38545
was found to be a close binary system (see Marchetti et al. 2001 for
all references). However, Faraggiana et al. (2001) were not able to
decide if the found ``shell'' features (see Section~\ref{gas}) are due to
an apparent spectroscopic binary nature or due to circumstellar gas. The
same is true for HD~111786 for which Faraggiana et al. (1997) reported
a spectroscopic binary nature based on IUE data. Later, Faraggiana et al. (2001)
speculated that a ``clump'' of at least five objects mimics one spectrum.
Since we have chosen to present only well established \LB objects, the 
already published results for HD~38545 and HD~111786 will be mentioned in the
following Sections but they are not included in our sample of investigated objects.

\subsubsection{Dust around \LB stars} \label{dust}

Sadakane \& Nishida (1986) found an infrared excess for two 
\LB stars, namely HD\,31295 and HD\,125162, from a 
cross correlation of the IRAS Faint Source Catalogue and the Bright 
Star Catalogue. King (1994) searched the IRAS catalogues for infrared detections
of \LB stars from the catalogue by Renson et al. (1990).
He found that only two well established \LB stars (HD\,31295 and HD\,125162) show an
excess in at least one IRAS band. Moreover he concluded that the minimum 
gas mass necessary to cause the peculiar abundance 
pattern in \LB stars is so low, that the associated dust 
mass $\sim 10^{-7}$~M$_\odot$ may even be below the detection limit of 
IRAS or submillimeter telescopes.

Cheng et al. (1992) report the detection of an infrared excess for
the \LB star HD\,110411 based on the IRAS Faint Source 
Catalogue. The colours of the excess are similar to those of Vega.
Note that King (1994) did not confirm
the infrared excess for HD\,110411.

Using ISOPHOT data from the Infrared Space Observatory (ISO; Kessler et al. 1996) 
Fajardo-Acosta et al. (1999) confirmed the infrared excess of HD\,110411 and report
a tentative detection of HD\,192640 (only at 20~$\mu$m).

The infrared excess for Vega, an object closely connected to the
\LB group, has been analyzed in more 
detail. The disk as seen by IRAS has a FWHM of 30$\arcsec$ (Aumann 1991)
corresponding to a disk radius of $\sim 115$~AU. Heinrichsen et al. (1998) 
observed Vega with the ISOPHOT instrument 
on board of the ISO satellite and found that the disk is resolved 
at 60 and 90~$\mu$m yielding a FWHM of 22 and 36$\arcsec$ respectively.
The disk is, as the star (Gulliver et al. 1994), seen pole-on.
Submillimeter images of Vega reveal an offset of 9$\arcsec$ with respect
to the stars position and a FWHM of $24\times 21\arcsec$ (Holland et al. 1998). 
Assuming typical grain sizes for fitting the infrared excess,
Ciardi et al. (2001) using the Palomar Testbed Interferometer found 
tentative evidence for a disk inside 4~AU contributing about 5~\% of 
the flux at 2.2~$\mu$m.

It is still difficult to answer the question whether circumstellar
dust is a basic characteristic of \LB stars or just
as frequent or rare as for all other A-type stars. The general
problem is that in most cases the detectors lack the sensitivity 
to detect the stellar atmospheres at infrared wavelength. This makes 
it difficult to exclude an infrared excess in the case of a non-detection.

Habing et al. (2001) overcame this problem by choosing bright
stars for which even the photosphere was detectable with the 
ISOPHOT instrument at 60~$\mu$m. Hence a pure photospheric
detection ruled out any dust infrared excess. In this way they
concluded that 6 out of their 15 A-type stars have circumstellar
dust and that these are mostly the younger ones between 200
and 400~Myr (post-main-sequence tracks were used to determine 
the stellar age of the A-type stars). The general conclusion
is that most stars arrive at the main-sequence surrounded by
a disk which then disappears within about 400~Myr.

Dunkin et al. (1997) and Kamp et al. (2002) studied the abundance
pattern in a sample of dusty A stars. Their results clearly show that
the presence of dust around A-type stars
does not necessarily imply the presence of the typical \LB
abundance pattern.

\begin{figure*}
\begin{center}
\resizebox{\hsize}{230mm}{\includegraphics{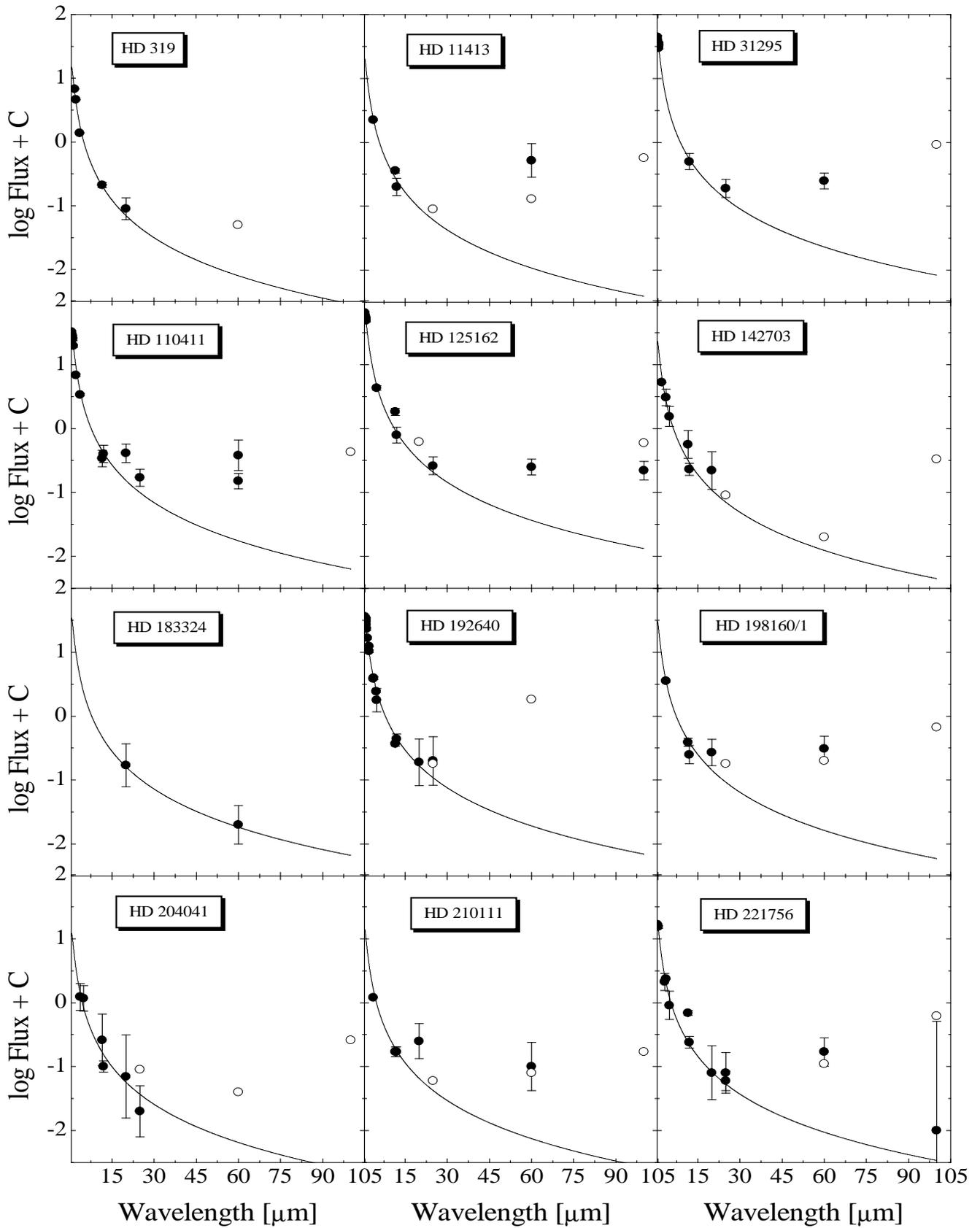}}
\caption[]{Fluxes taken from Table\,\ref{result}
for HD~319, HD~11413, HD~31295, HD~110411, HD~125162,
HD~142703, HD~183324, HD~192640, HD~198160/1, 
HD~204041, HD~210111 and HD~221756 with the blackbody
curve $T_{\rm eff}$\,=\,8500\,K; open circles are upper limits.}
\label{ir_f1}
\end{center}
\end{figure*}

\begin{figure*}
\begin{center}
\resizebox{\hsize}{230mm}{\includegraphics{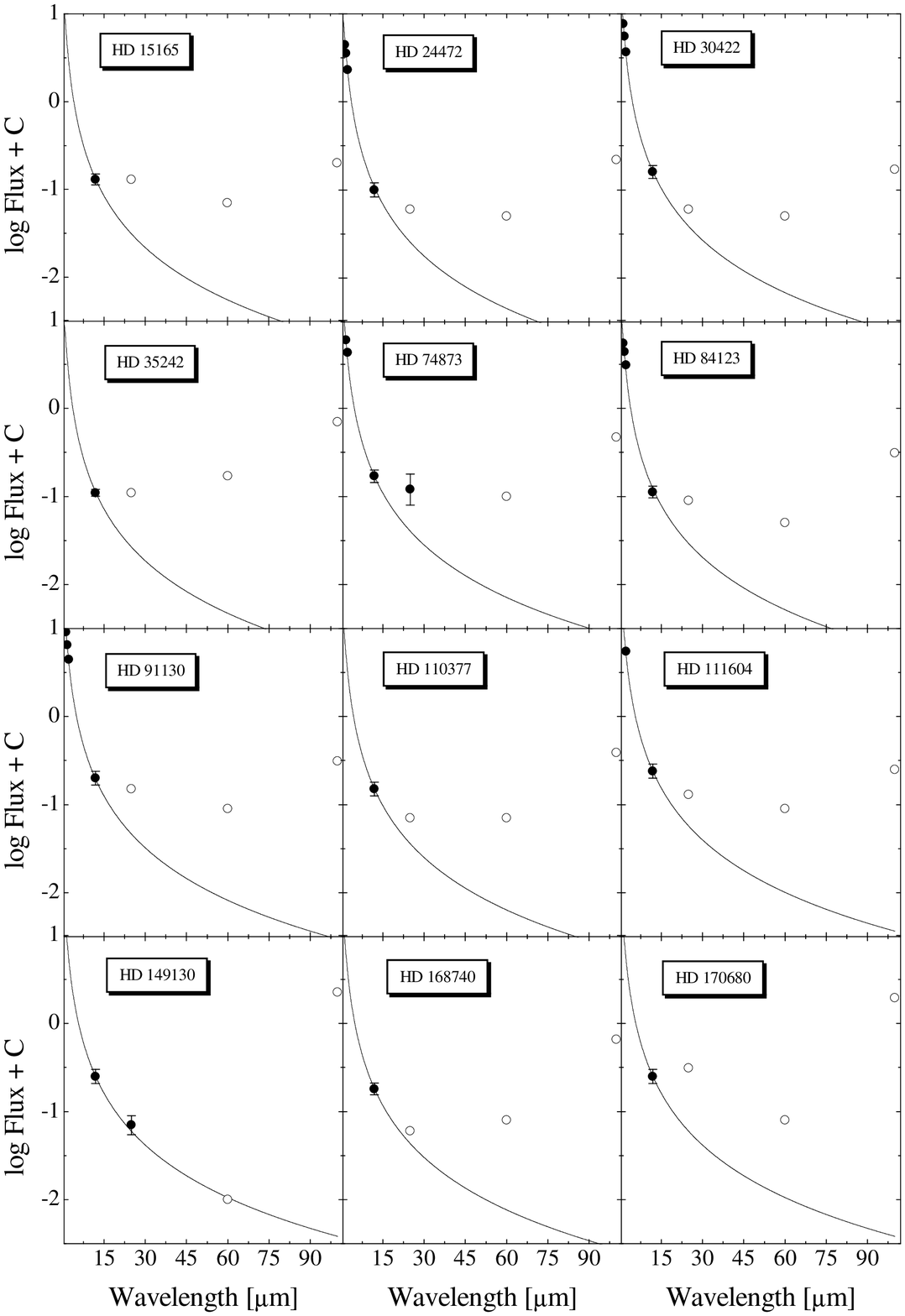}}
\caption[]{Fluxes taken from Table\,\ref{result} 
for HD~15165, HD~24472, HD~30422, HD~35242, HD~74873, HD~84123,
HD~91130, HD~110377, HD~111604, HD~149130, HD~168740 and HD~170680
with the blackbody
curve $T_{\rm eff}$\,=\,8500\,K; open circles are upper limits.}
\label{ir_f2}
\end{center}
\end{figure*}

\subsection{Gas around \LB stars} \label{gas}

If the inclination of a circumstellar disk is favourable or
if a star is surrounded by a shell, narrow absorption cores
can be observed on top of the stellar spectrum. These arise
from the additional absorption due to circumstellar material.
Similar narrow absorption lines can form in the interstellar
medium between the observer and the star. The most commonly
observed lines are Ca\,{\sc ii}\,K and Na\,{\sc i}\,D.

There are different criteria to distinguish between
circumstellar and interstellar origin. The doublet line
ratio of Na\, for example is a measure of the
optical thickness of the absorbing medium. If the lines
are optically thin, the ratio is 2.0, while it decreases
to 1.0 in the case of fully saturated lines (see e.g.
Spitzer 1968). Moreover an equivalent width ratio 
W(Ca\,{\sc ii})/W(Na\,{\sc i}) much larger than 1.0 points 
towards a circumstellar origin (Lagrange-Henri et al. 1990).

However, the detection of narrow gaseous circumstellar or interstellar lines
(often denoted as ``shell'' signs) 
seriously questions the spectral classification criteria of the \LB
group as established by Gray (1988).
He {\it explicitly excluded} classical shell stars from the group of \LB stars. 
Let us recall the definition of a classical shell star (Jaschek \& Jaschek 1987):
a B/A type shell star is characterized by the simultaneous presence of 1) broad absorption
lines and 2) sharp absorption lines which arise from ground states or metastable levels.
The number of lines often suggests the spectrum of a supergiant with an apparent weakness
of the Mg\,{\sc ii}\,4481 and Si\,{\sc ii}\,4128-30 lines. But the hydrogen and helium
lines clearly point at a luminosity class V classification. Abt \& Moyd (1973) and 
Jaschek et al. (1988)
found that the strength of shell lines (mostly Ca\,{\sc ii}, Ti\,{\sc ii}, Fe\,{\sc ii} 
and Sc\,{\sc ii}) vary in time. Usually, sharp absorption features were detected
in Ca\,{\sc ii}\,K, but only marginal absorption features in the hydrogen lines. 
The latter are exactly
the characteristics detected for members of the \LB group (Section~\ref{obs_gas}). But
Heiter (2002) showed that the classical shell stars do not share the typical abundance
pattern of the \LB group. The detection of shell features does, therefore, {\it not
a-priori exclude} an object as being member of the \LB group.

\subsubsection{Confrontation with observations} \label{obs_gas}

{\it HD~11413} and {\it HD\,198160/1:}
Holweger \& Rentzsch-Holm (1995) found narrow absorption
cores in Ca\,{\sc ii}\,K for these objects. 
The presence of these features is correlated with stellar properties like gravity and
rotational velocity indicating a circumstellar rather 
than interstellar origin. \\
{\it HD~38545:}
Bohlender \& Walker (1994) report the detection of a
circumstellar shell around HD\,38545 from narrow absorption
features in Fe\,{\sc ii}, Ti\,{\sc ii}, Ca\,{\sc ii} and
the Balmer lines.
Hauck et al. (1995, 1998) found evidence for circumstellar
shell features in the Ca\,{\sc ii}\,K line for nine candidate
\LB stars and they confirm the circumstellar shell. 
They use Crutcher's formula (Crutcher 1982) to derive
the velocity component of the ISM in the direction of the
star and conclude that their features are clearly
circumstellar.
Grady et al. (1996) confirmed the detection of shell
lines in the spectrum and announced
the first detection of accreting circumstellar gas in a
\LB star. They observed redshifted narrow
absorption components of Fe\,{\sc ii} and Zn\,{\sc ii}
in the mid-ultraviolet resembling those regularly observed
in the spectrum of $\beta$~Pictoris. 
Kamp et al. (2002) noted the variability of the 
narrow absorption feature as well as
an equivalent width ratio W(Ca\,{\sc ii})/W(Na\,{\sc i})
below unity. The variability underlines the presence of circumstellar
gas. McAlister et al. (1993) and Marchetti et al. (2001) reported the
detection of a close companion of HD~38545 using speckle interferometry.\\
{\it HD\,39283, HD\,84948,
HD\,98353, HD~125162, HD~183324} and {\it HD\,217782:}
Andrillat et al. (1995) examined a sample of 
20 candidate \LB stars in the near-IR spectral region
and found that HD\,39283, HD\,84948,
HD\,98353, and HD\,217782 exhibit evidence of
circumstellar shells. The spectra show clear 
signatures of shells as seen in classical A-type shell stars.
Three objects (HD\,39283, HD\,98353 and HD\,217782) have been 
excluded from the \LB group because they do not share the
typical characteristics (Gray \& Garrison 1987, Faraggiana et al. 1990).
HD~84948 was later found to be a spectroscopic binary system
(Iliev et al. 2002). The shell evidence for HD\,125162 and HD\,183324 is still
dubious. Note that Bohlender et al. (1999) do not confirm the detection of 
circumstellar gas in HD\,183324.\\
{\it HD~111786:} 
Gray (1988) was the first to notice a weak narrow absorption
core in the Ca\,{\sc ii}\,K line of HD\,111786, which he
suspected to be of circumstellar origin. 
Holweger \& St\"{u}renburg (1991) report the presence of
narrow absorption components in the Na\,{\sc i}\,D and 
Ca\,{\sc ii}\,K lines. The ratio of 
the two Na\,{\sc i}\,D lines is found to be $\sim 1.3$, hence a larger 
optical depth. This star turned out to be in fact a spectroscopic binary consisting 
of a broad-lined \LB star and a narrow-lined 
F-type star (Farragiana et al. 1997). Holweger et al. (1999) 
Ca\,{\sc ii}\,K. They argue that the Ca\,K feature
seen can not be produced by the binary nature
of this star, that is by the superposition of a \LB star and 
an early F star. Moreover the strong variability of this
feature seems to eliminate an interstellar origin. 
For HD\,111786, the equivalent width ratio 
W(Ca\,{\sc ii})/W(Na\,{\sc i}) could be determined and 
was found to be about unity. The equivalent width criterium should not be taken too
strict, because a range of different ionisation conditions or 
elemental abundances in the circumstellar gas leads to deviations
from W(CaII)/W(NaI) $\gg$ 1. \\
{\it HD\,142994, HD\,192640} and {\it HD\,221756:}
Bohlender et al. (1999) reported the possible detection
of strong circumstellar Na\,{\sc i}\,D features in
the spectra of HD\,142994 and HD\,221756. Moreover they note 
the presence of a most likely interstellar feature in HD\,192640. \\
{\it HD\,193256} and {\it HD\,193281:}
Holweger \& St\"{u}renburg (1991) report the presence of
narrow absorption components in the Na\,{\sc i}\,D and 
Ca\,{\sc ii}\,K lines for HD\,193256 and HD\,193281. The ratio of 
the two Na\,{\sc i}\,D lines is found to be $\sim 2.0$, a possible indication 
for their interstellar origin. The results for HD\,193256 was
later confirmed by Holweger \& Rentzsch-Holm (1995). Bohlender et al. (1999)
cautioned against a circumstellar origin of the narrow absorption
features in HD\,193256 and HD\,193281.

To summarize, there is definitely evidence for circumstellar gas around
some \LB stars, but not around all. Moreover the presence of
gas {\em and} dust is only observed in one \LB star unambiguously,
HD\,221756. Table~\ref{lb_cs} gives an overview of the detection of
dust and/or gas around \LB stars. If the material is confined in a 
disk coplanar with the star, only a fraction of the stars will 
show narrow absorption features and/or shell characteristics, 
namely those, where the star is seen nearly equator-on. On the other 
hand at least the more distant \LB stars beyond the 
Local Bubble, $d > 100$~pc, are expected to show interstellar 
absorption features.

\section{Overall strategy and general remarks} \label{strat}

The well established \LB\ stars were taken from the list of 
Gray \& Corbally (1998) and Paunzen (2001) with the exception of apparent 
spectroscopic binary systems (e.g. HD~64491, HD~141851 and HD~148638). 
In total, the list comprises of 65 objects. We have
explicitly rejected any spurious objects. King (1994) has used the
first catalogue of \LB candidates from Renson et al. (1990) for his target
selection including more than 100 objects. At least 50\,\% of this sample was
already recognized as being misclassified and several new
\LB type objects were discovered since then (Faraggiana \& Bonifacio 1999, Paunzen
2001). A one-to-one comparison with the work of King (1994) is therefore
not straightforward. 

The literature was searched for measurements of well established \LB
type objects red-ward of 7000\AA\,(approximate Johnson $R$). A
blackbody radiation curve follows a standard ATLAS9 model very well
within this spectral region.
The {\it same} blackbody radiation curve for a temperature of 8500\,K
was used for all program stars. This curve is shifted by a constant for
the individual objects mainly due to the different distance and 
interstellar absorption and thus reddening.

All measurements were transformed to Jansky with the calibrations listed
in the following sections. The results are listed in Table~\ref{result}
where {\it italized} values are only {\it upper limits}.

Figures\,\ref{ir_f1}\,and\,\ref{ir_f2} show the fluxes for all objects
with photometric measurements beyond 2.2\,$\mu$m (open circles denote
upper limits) together with the blackbody radiation curves. 

\subsection{Measurements up to 1.1\,$\mu$m}

For this work we have used two different photometric systems (13 color
and Johnson $RI$) taken from the General Catalogue of Photometric Data (GCPD;
http://obswww.unige.ch/gcpd/) as well as spectrophotometry given by Oke (1967).
This data is not color corrected because all calibrations
are derived relative to an A0\,V star (or Vega) with an effective
temperature close to our program stars. The listed (e.g. Johnson \& Mitchell
1975) color corrections are indeed close to zero. The observed magnitudes were
directly converted into fluxes using the calibrations by Johnson \& Mitchell
(1975; 13 color system), Beckwith et al. (1976; Johnson $RI$) and 
Oke \& Gunn (1983; spectrophotometry).

The absolute values sometimes differ which is probably caused by the contribution
of Paschen lines in the relevant spectral domain.

\subsection{Johnson $JHKLM$ photometry}

Most of the data are taken from the 2 micron survey (Skrutskie et al. 1997) which lists
$JHK$ colors and their calibration. The other data (Baruch et al. 1983, Oudmaijer et al.
2001, GCPD)
were calibrated using the values from Bessell \&
Brett (1988). Both sources take an A0\,V star as standard. The measurements follow
the blackbody radiation curve very well.

Gerbaldi (1991) presented $JHK$ photometry for 25 candidate \LB type objects. 
Unfortunately, instead of individual photometric magnitudes, calibrated
spectral types are listed derived from $(V-K)$ and $(J-K)$ values. Hence, we were
not able to include this data.

\subsection{IRAS photometry}

The results of the IRAS satellite have been extensively analyzed in the
past (Cheng et al. 1992, King 1994). The data is extracted from the IRAS Point
Source Catalogue Version 2 and Faint Source Catalogue. 
Measurements with positional offsets more than 3$\sigma$ from
the optical source were neglected. Most of the data for 25, 60 and 100\,$\mu$m
are only upper limits (King 1994) and determined by the high background within
these passbands. Since we want to get ``upper limits'' for possible
disks around these objects, an independent estimation of the expected background
was performed. Schlegel et al. (1998) list maps of dust infrared emission for
the 100\,$\mu$m band derived from the COBE/DIRBE and IRAS/ISSA measurements. 
This consortium also presents maps for 60$\mu$m (http://astron.berkeley.edu/dust/).
These maps are used to derive a predicted background signal for a given galactical
longitude and latitude which is then subtracted from the IRAS 60 and 100\,$\mu$m measurements. 
The remaining upper limits for the 60\,$\mu$m band are typically less than 0.1\,Jansky 
whereas the values for the 100\,$\mu$m band are slightly higher. Such corrected values
are still upper limits and not apparent detections of fluxes from the targets themselves.

As a last step, the fluxes
are color corrected using a 10000\,K blackbody radiation (IRAS Explanatory Supplement VI.C.6). 
This has to be done because the absolute fluxes are not derived by using an A0\,V standard 
star but by an internal calibration lamp (as for ISO).

\begin{table}
\caption{List of observations performed with the ISO-satellite for
well established members of the \LB group; the column ``mode'' denotes
chopped (c) or staring (s) observations.} 
\label{iso_log}
\begin{tabular}{lcccc}
\hline
\multicolumn{1}{c}{HD} & Inst. & TDT\,number & Program & mode \\
& \\
\hline
319 & PHT03 & 37501221 & RSTENCEL & s \\
    & PHT22 & 37501238 & RSTENCEL & s \\
11413 & PHT03 & 35302022 & RSTENCEL & s \\
      & PHT22 & 35302039 & RSTENCEL & s \\
110411 & PHT03 & 20201523 & RSTENCEL & s \\
       & PHT22 & 20201540 & RSTENCEL & s \\
125162 & PHT03 & 16500415 & CWAELKEN & c \\
	   & PHT03 & 35101427 & WWEISS & c \\
       & SWS01 & 35101303 & WWEISS & c \\
142703 & PHT03 & 43100804 & WWEISS & c \\
142994 & PHT03 & 09000784 & RSTENCEL & s \\
183324 & PHT03 & 12800785 & RSTENCEL & s \\
       & PHT22 & 53400928 & RSTENCEL & s \\
192640 & PHT03 & 34700540 & RSTENCEL & s \\
       & PHT22 & 34700537 & RSTENCEL & s \\
	   & PHT03 & 38101529 & WWEISS & c \\
       & SWS01 & 38101302 & WWEISS & c \\
       & SWS01 & 38101406 & WWEISS & c \\
193256 & PHT03 & 36902530 & RSTENCEL & s \\
       & PHT22 & 36902547 & RSTENCEL & s \\
198160 & PHT03 & 54601631 & RSTENCEL & s \\
       & PHT22 & 54601648 & RSTENCEL & s \\
204041 & PHT03 & 71101211 & WWEISS & c \\
210111 & PHT03 & 36902628 & RSTENCEL & s \\
       & PHT22 & 36902645 & RSTENCEL & s \\ 
221756 & PHT03 & 38201729 & RSTENCEL & s \\
	   & PHT03 & 41701332 & WWEISS & c \\
       & PHT22 & 38201746 & RSTENCEL & s \\     
\hline
\end{tabular}
\end{table}

\subsection{ISO photometry}

In this section we present the multifilter photometric
observations for \LB stars from the ISO satellite. The results of our program
(WWEISS\_LBODISK) and the group of C.~Waelkens (one measurement for
HD~125162; CWAELKEN\_VEGASTAR) have not been published before. Fajardo-Acosta et al. (1999)
presented the results of their working group (RSTENCEL\_VEGADIS7) which was partly
focussed on the group of \LB stars. They only list fluxes for HD~11413, HD~110411
and HD~192640 which were reduced with the PIA software (V6.3).

There are two different modes of ISO observations:
\begin{itemize}
\item chopped observations: the target and sky background
is measured four times separately.
\item staring observations: the measurements are made once at the
target and once at the sky background. The ISO consortium soon realized that
this mode was not very stable.
\end{itemize}
However, we treat both modes as having equal quality because we are not able to give
any reasonable weights based on physical grounds. All data were newly reduced 
with the PIA software (V9.1; Gabriel et al. 1997) which should yield better 
calibrations than before. 
The color correction was done according to Chapter C.3 of the ISOPHOT/PIA manual.
Fajardo-Acosta et al. (1999)
give a calibration of ISO data at 3.6, 11.5 and 20\,$\mu$m in comparison
with IRAS and ground-based measurements. We have checked these calibrations and found
an excellent agreement with the data reduced with the new software. We also have
to emphasize that no study for the validity of the calibration for 
the measurements of A-type stars in the 25, 60
and 100\,$\mu$m passbands exist. For the few cases for which IRAS and ISO measurements
in the relevant passbands are available, they are in excellent agreement. We are therefore
confident that the results from the PIA software (V9.1) have at least the same quality as the
corresponding IRAS measurements. Table~\ref{iso_log} lists all used multifilter
photometric as well as spectroscopic (Section~\ref{iso_sws01}) ISO observations.

\begin{table}
\caption{List of identified H\,{\sc i} lines for the ISO-SWS spectra of
HD~125162 and HD~192640 (Fig.~\ref{sws_spec}) with the transition i$\rightarrow$j} 
\label{sws_h}
\begin{tabular}{ccr|ccr|ccr}
\hline
$\lambda$ & i & \multicolumn{1}{c|}{j} & $\lambda$ & i & 
\multicolumn{1}{c|}{j} & $\lambda$ & i & \multicolumn{1}{c}{j} \\
$[\mu\rm m]$ & & & $[\mu\rm m]$ & & & $[\mu\rm m]$ & & \\
\hline
2.526 & 5 & 16 & 2.675 & 5 & 13 & 3.297 & 5 & 9 \\
2.564 & 5 & 15 & 2.758 & 5 & 12 & 3.741 & 5 & 8 \\
2.613 & 5 & 14 & 2.873 & 5 & 11 & 3.749 & 6 & 17 \\
2.626 & 4 & 6 & 3.039 & 5 & 10 & 3.819 & 6 & 16 \\
\hline
\end{tabular}
\end{table}

\begin{figure*}[t]
\begin{center}
\resizebox{\hsize}{110mm}{\includegraphics{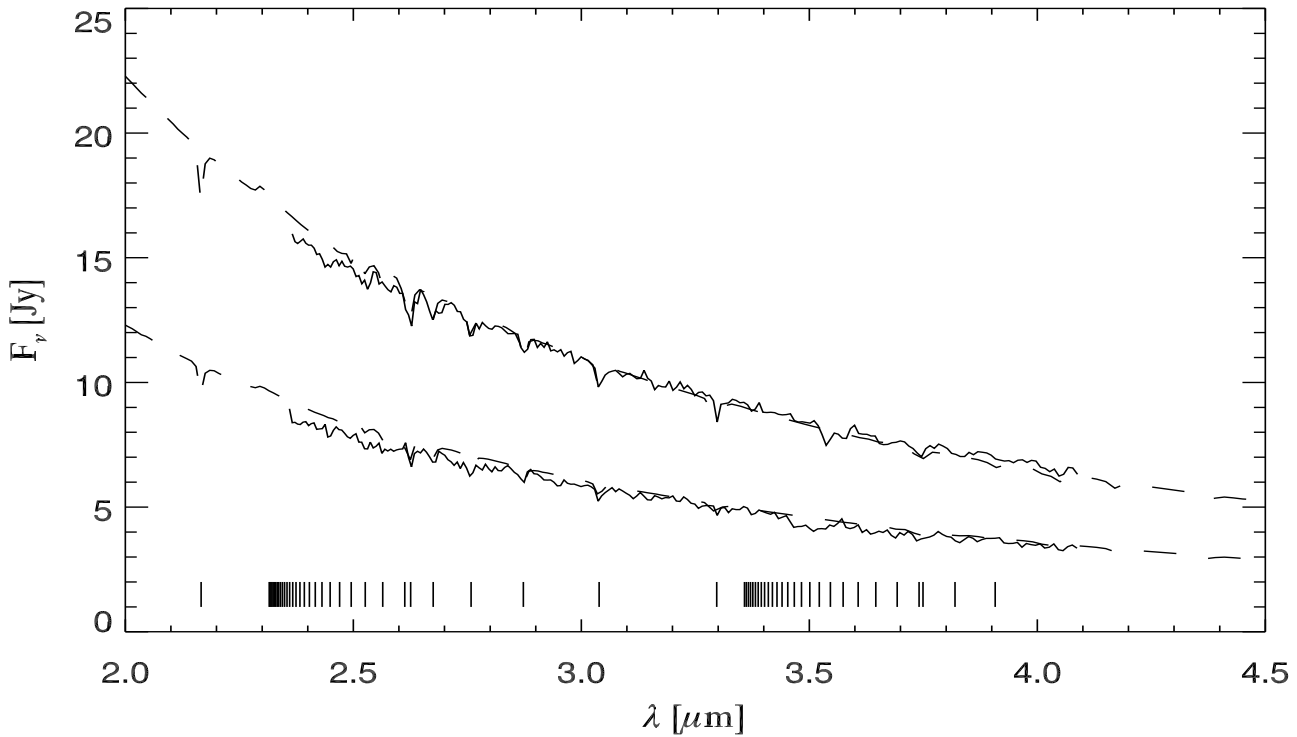}}
\caption{ISO-SWS spectra for HD~125162 (upper
lines) and HD~192640 (lower lines) compared to respective
models of the stellar photosphere (ATLAS9). The 
H\,{\sc i} lines are noted at the bottom.}
\label{sws_spec}		
\end{center}
\end{figure*}

\subsection{ISO-SWS spectroscopy for HD~125162 and HD~192640} \label{iso_sws01}

The ISO-SWS spectroscopy is well suited to find atomic
emission lines of H\,{\sc i}, and molecular lines, like H$_{\rm 2}$
or CO. This would allow to put further constraints on
the accretion scenario and the chemical composition of
the circumstellar material.

SWS spectroscopy was carried out for two well established
\LB stars, the prototype HD~125162 and HD~192640. 
The observation log is listed in Table\,\ref{iso_log}. The
reductions are performed using the standard software as provided
by the ISO-consortium. The two individual
spectra for HD~192640 were added.
The signal-to-noise ratio of the spectra
is about 15 to 20. For the wavelength region redder than 4\,$\mu$m
only upper limits due to the background noise can be given. 

\begin{figure}
\begin{center}
\resizebox{\hsize}{!}{\includegraphics{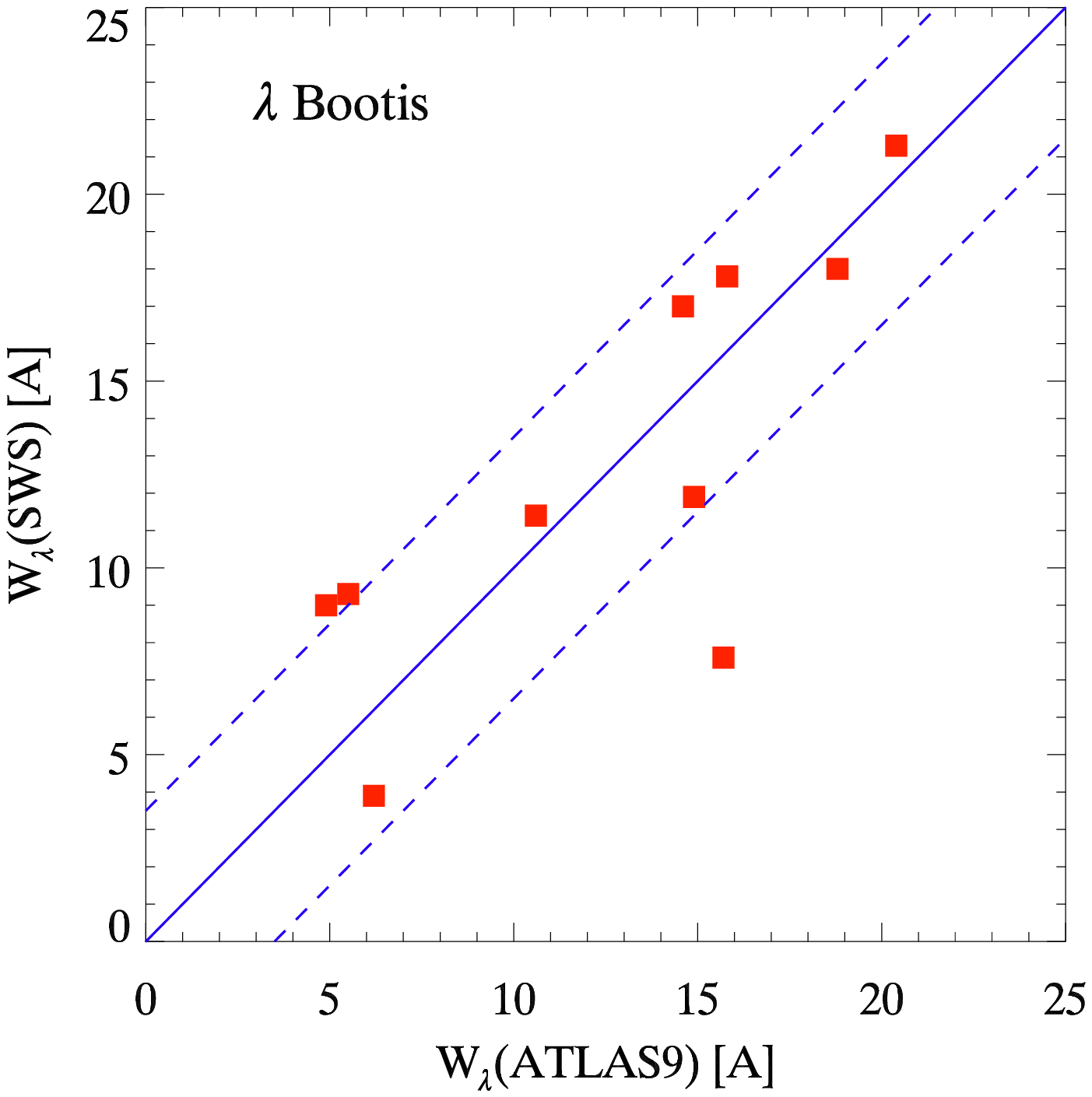}}
\resizebox{\hsize}{!}{\includegraphics{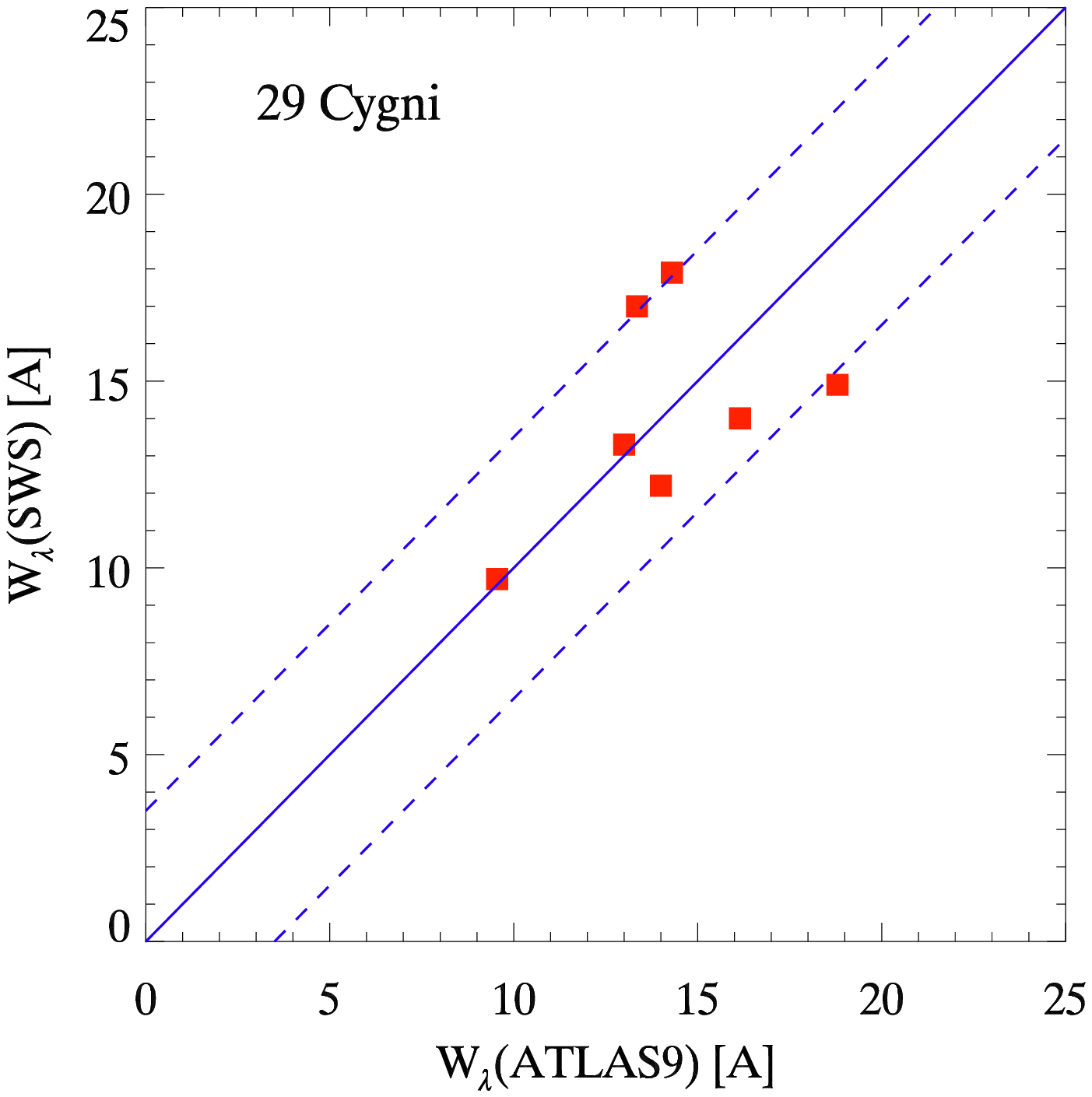}}
\caption{Comparison of the equivalent widths of the H\,{\sc i} lines
from ATLAS9 models and the observed SWS spectra for HD\,125162 (upper
panel) and HD\,192640 (lower panel). The dotted lines indicate the error
due to the low signal-to-noise ratio and difficulties due to
the normalization.}
\label{sws_eq}		
\end{center}
\end{figure}

The absolute calibration of the spectra is very good. 
Figure~\ref{sws_spec} shows a comparison
with theoretical ATLAS9 models (Kurucz 1992) using the following stellar
parameters and distances (St\"{u}renburg 1993, Heiter et al. 1998,
Paunzen et al. 2002): $T_{\rm eff}$\,=\,9000\,K, log\,$g$\,=\,4.0\,dex,
$d$\,=\,30\,pc for HD~125162 and $T_{\rm eff}$\,=\,8000\,K, log\,$g$\,=\,4.0\,dex,
$d$\,=\,41\,pc for HD~192640. Because of the rather low signal-to-noise ratios
of the spectra, an error of $\pm$250\,K for the effective temperature
and $\pm$0.15\,dex for the surface gravity does not affect the conclusions.

From the spectra themselves, several H\,{\sc i} lines (mainly the 
Pfund series) are identified. Table~\ref{sws_h} lists the
identified lines which are noted at the bottom of Fig.~\ref{sws_spec}.
The theoretical spectra were smoothed to a resolution corresponding
to the SWS spectra in order to measure the equivalent widths 
of the hydrogen lines. Assuming a typical error of 1 and 
2.5\,\AA\,(taking also into account the difficulties of normalization)
for the synthetic and the observed spectra respectively, no emission
was found for the two \LB stars (Fig.~\ref{sws_eq}) within the
error limits. This means that the observed H\,{\sc i} lines are of pure
stellar origin ruling out an active accretion disk (as
found for some Herbig Ae/Be stars) around these objects. 

%The only unidentified strong feature found in these spectra is
%around 3.5\,$\mu$m. This feature is present in both overlapping
%orders of the spectrograph which seems to rule out an instrumental
%origin. Its equivalent width is 30\,\AA\,(HD~125162) and 
%70\,\AA\,(HD~192640) respectively. These values are just 
%approximations since the onset of the H\,{\sc i} series (n\,=\,6)
%causes difficulties in placing the continuum around 3.35\,$\mu$m.
%Figure~\ref{sws_35} shows the spectra of both program stars around
%3.5\,$\mu$m. A further identification is not possible up to now.

%\begin{figure}
%\begin{center}
%\resizebox{\hsize}{!}{\includegraphics{fig5a.eps}}
%\resizebox{\hsize}{!}{\includegraphics{fig5b.eps}}
%\caption{Observed (dotted lines) and
%synthetic (full lines) spectra around 3.5\,$\mu$m for HD~125162 (upper
%panel) and HD~192640 (lower panel).}
%\label{sws_35}		
%\end{center}
%\end{figure}

\begin{figure}
\begin{center}
\resizebox{\hsize}{150mm}{\includegraphics{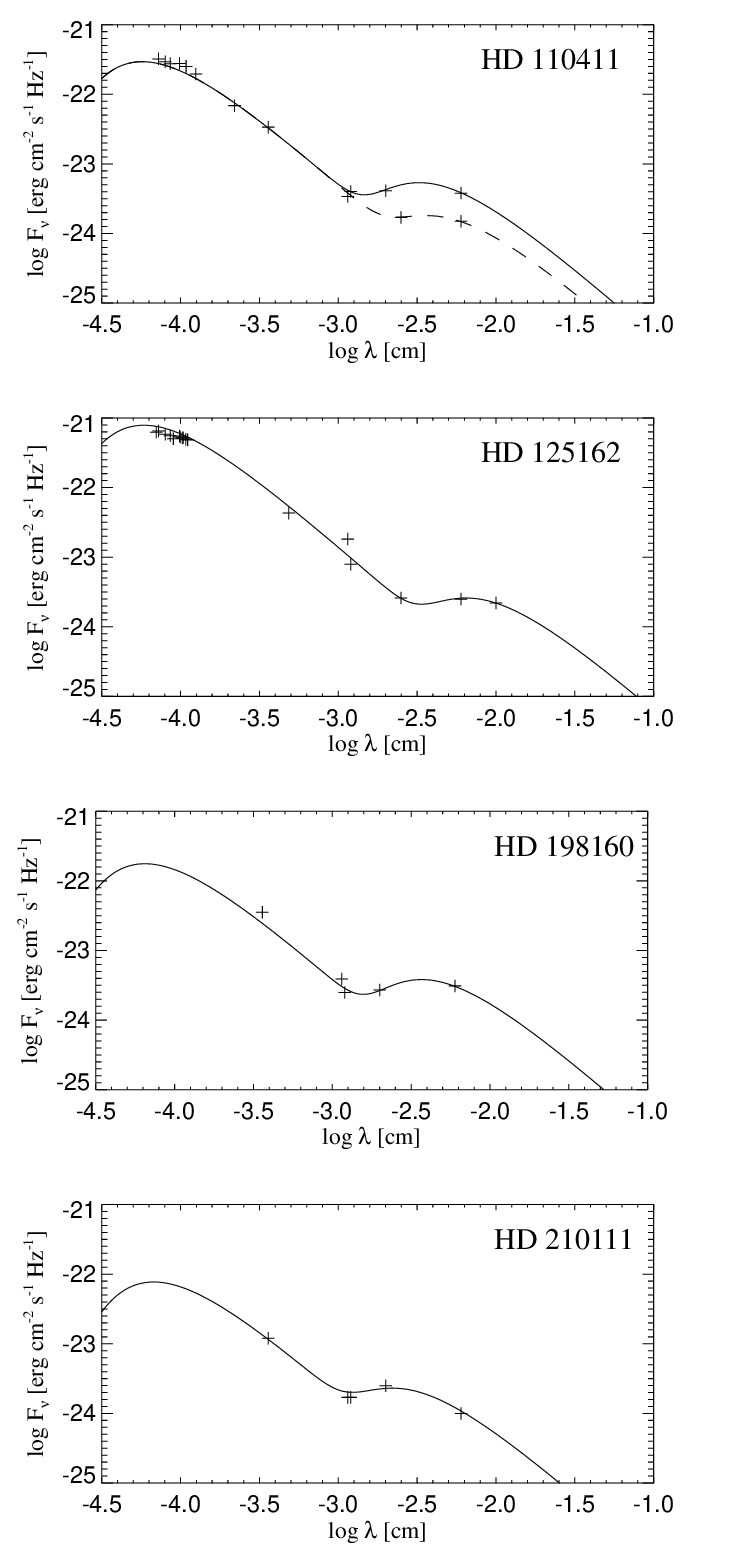}}
\caption{Spectral energy distribution for well established \LB stars with a
detected infrared excess; the values for the latter were taken from Table\,\ref{tab:ir}.
The full line for HD~110411 fits the ISO data whereas the dotted line
fits the IRAS data.}
\label{ex_plot}		
\end{center}
\end{figure}
\begin{table*}
\caption{Stellar parameters and characteristics of the infrared
excess: effective temperature T$_{\rm eff}$, stellar luminosity
$L_\ast/{\rm L}_\odot$, stellar radius $R_\ast/{\rm R}_\odot$ and distance
$d$ are taken from Paunzen et al. (2002) and the spectral type from
Gray \& Corbally (1993). The characteristic
dust temperature T$_{\rm dust}$ and fractional dust luminosity 
$L_{\rm IR}/L_\ast$ are deduced from the fits, while the
radiative equilibrium distance $r_{\rm dust}$ is derived from
Eq.(\ref{eq:rdust}).}
\begin{tabular}{llllllrrlr}
\hline\\[-2mm]
    HD    & Sp. Type & $T_{\rm eff}$/K & $L_\ast/{\rm L}_\odot$  & $R_\ast/{\rm R}_\odot$ &   $d$/pc   & 
            $T_{\rm dust}$  & $r_{\rm dust}/{\rm AU}$ & $L_{\rm IR}/L_\ast$  & 
            $r_{\rm dust}^{\rm ISM}/{\rm AU}$\\[2mm]
\hline\\
110411$^{\rm (a)}$ & A0\,Va ($\lambda$ Boo)   & 8930   &  16.59                  & 1.71                   & 37.0     &
            145             &  11.3                   & $1.7 \times 10^{-4}$ &
 330\\
110411$^{\rm (b)}$ & A0\,Va ($\lambda$ Boo)  & 8930   &  16.59                  & 1.71                   & 37.0     &
            116             &  17.7                   & $4.2 \times 10^{-5}$ &
 609\\
125162    & A0\,Va $\lambda$ Boo   & 8720            &  19.05                  & 1.92                   & 30.0     &
             70             &  51.3                   & $2.2 \times 10^{-5}$ &
 2570\\
198160    & A2\,Vann wk $\lambda$4481 & 7870            &  22.91                  & 2.58                   & 73.0     &
            130             &  16.0                   & $3.2 \times 10^{-4}$ &
 476\\
210111    & kA2hA7mA2\,Vas $\lambda$ Boo & 7550            &  16.98                  & 2.41                   & 79.0     & 
            200             &   6.0                   & $4.3 \times 10^{-4}$ &
 121\\[3mm]
 \hline
 \\
\multicolumn{5}{l}{(a) fit to ISO fluxes}\\
\multicolumn{5}{l}{(b) fit to IRAS fluxes}\\
\end{tabular}
\label{tab:ir}
\end{table*}

\subsection{(Sub)millimeter observations at the Heinrich-Hertz-Telescope (HHT)}

We have also observed five members of the \LB group at 
(sub)millimeter wavelengths, in order to find signs of cold dust. The 
CO\,$(2-1)$ line at 238.538\,GHz was chosen, because it traces the molecular 
gas on excitation conditions characteristic for both the circumstellar 
and interstellar environment. Rentzsch-Holm et al. (1998) show that CO is not 
photodissociated in the circumstellar environment of stars with effective 
temperatures below 9000 K. We have selected five well established \LB 
stars with effective temperatures below 9000\,K where signs of gas 
and/or dust have already been detected: HD 31295, HD 74873, HD 125162, 
HD 192640, and HD 221756. 

The observations were made at the Heinrich-Hertz-Telescope (Baars et al. 1999)
on Mt. Graham, Arizona, in November 2000. We used the MPIfR 19 element 
bolometer array, that is sensitive at a spectral-response weighted frequency 
of 347\,GHz with 110\,GHz bandwidth. The array elements were calibrated 
by observations of the planets (Mars and Uranus were used as primary calibration
sources).
For the spectral line observations, a facility 215\,$-$\,275\,GHz SIS receiver from Steward 
Observatory was used. 
Spectra were derived using a filterbank with 250\,kHz channel spacing.
The spectral line observations achieved a mean system temperature of 
356\,K (ranging between 303 and 435\,K). For both line and continuum observations, 
the wobbler was used to 
remove the atmospheric emission, with 100$\arcsec$ and 200$\arcsec$ beam throw
(the latter corresponds to the maximum beam separation of the array's 
elements), respectively. The beam sizes are 31$\arcsec$ for the CO 
spectroscopy and 21$\arcsec$ for the 347\,GHz photometry. 

The results are summarized in Table~\ref{hht}. No source was detected in the 
continuum, only one in the CO\,$(2-1)$ line, which origin is unclear. Here
are the results for the individual objects in more details:
\\[1.5ex]
\noindent
{\it HD 31295} and {\it HD 74873}: Only 347\,GHz photometry, but no detection.
{\it HD 125162}: The source was observed in the CO line, but not detected. \\
{\it HD 192640}: The source was observed in the 347\,GHz continuum and the 
CO line. The continuum of the source was not detected. A Gaussian fit 
to the CO line yields a system velocity of 4.36\,km\,s$^{-1}$ and a linewidth of 
1.66\,km\,s$^{-1}$ (FWHM). However, a contamination by emission in the 
off-beam at $-5.2$\,km\,s$^{-1}$ suggests extended CO emission of interstellar rather 
than circumstellar origin. This suggestion has been confirmed by an incomplete 
map peaking at offsets of $\Delta\alpha = -65\arcsec$ and $\Delta\delta = -152\arcsec$.
The CO flux and line area within the telescope's beam, at the source position,
are 11.4\,Jy and 20.2\,Jy\,km\,s$^{-1}$, respectively. This emission is dominated by 
extended interstellar material. Dame et al. (1987) and Uyaniker et al. (2001)
investigated survey data of the infrared, H\,{\sc i} and CO emission in the
relevant area. They conclude that the Orion local spiral arm is seen tangential towards the 
Cygnus region. This results in intense radio emission with a complex morphology. Discrete
CO structures are seen at velocities between $-10$ and $+10$\,km\,s$^{-1}$ whereas 
large scale diffuse CO emission exists at velocities
smaller than $-12$\,km\,s$^{-1}$. \\
{\it HD 221756}: Only CO spectroscopy, but no detection.

\begin{table}
\caption{Results of (sub)millimeter observations at the HHT.}
\label{hht}
\begin{tabular}{lcc}
\hline
HD & \multicolumn{2}{c}{point source sensitivity [Jy]} \\
   & CO(2-1) (238.5 GHz) & continuum (347 GHz) \\
\hline
125162 & 0.71 & $-$   \\
192640 & 1.87 & 0.017 \\
221756 & 2.41 & $-$   \\
31295  & $-$  & 0.013 \\
74873  & $-$  & 0.012 \\
\hline
\end{tabular}
\end{table}

\section{Results} \label{sect_res}

In the literature, three well established \LB stars with an apparent infrared excess
were reported: HD~31295, HD~110411 and HD~125162 (see
Section~\ref{res_lit}). Furthermore, a tentative detection for HD~11413 and HD~192640
was listed by Fajardo-Acosta et al. (1999). From our investigation, we
are able to confirm the previous findings for HD~31295, HD~110411 and HD~125162.
In addition, an infrared excess for HD~74873, HD~198160, HD~210111 and HD~221756
was found. This result is based on only one measurement at 60\,$\mu$m for  
HD~210111 and HD~221756 as well as 25\,$\mu$m for HD~74873
which makes a further investigation of the dust properties impossible.

We are not able to support the results for HD~11413 and HD~192640 as reported
by Fajardo-Acosta et al. (1999). The ISO measurements for HD~192640 at 20 and 25\,$\mu$m
are 0.19(13) and 0.20(28)\,Jy, respectively whereas the upper limit
from the IRAS photometry at 25\,$\mu$m is 0.18\,Jy (Table\,\ref{result}).
The ISO measurements for HD~11413 at 60\,$\mu$m of 0.52(18)\,Jy is not in line
with the upper limit of the IRAS value (0.13\,Jy).
This clearly calls for further measurements with higher sensitivity to
to check the presence of an infrared excess for both objects. 

Furthermore, we deduced the percentage of infrared measurements for
well established \LB stars redward of 20\,$\mu$m and the ratio of objects
with infrared excesses. Table\,\ref{result} includes 26 stars 
from which 11 have only upper limits, two are doubtful cases and
7 show no sign of an infrared excess. Hence, 23\,\% of the \LB stars 
show an infrared excess.

\subsection{Models of Infrared Excesses}

For four stars in our sample, namely HD\,110411, HD\,125162, 
HD\,198160 and HD\,210111, the infrared excess is detected
at two wavelengths. Any detailed modelling of the
excess emission making use of more than two free parameters
would be an overinterpretation of the data. Hence, we restrict
ourselves to simply use a black body temperature $T_{\rm dust}$ 
and solid angle $\Omega$ to fit the infrared excess of these 
four stars
\begin{equation}
F_{\rm dust} = B_\nu(T_{\rm dust}) \Omega 
               ~~~{\rm erg}~{\rm cm}^{-2}~{\rm s}^{-1}~{\rm Hz}^{-1}~.
\end{equation}
The total flux is given by adding the stellar flux and the
infrared excess flux. From this fit, we obtain a characteristic
dust temperature. Assuming large black body dust grains and
radiative equilibrium, we can obtain the distance $r_{\rm dust}$ 
from the star at which the dust is heated to this temperature 
(Backman \& Paresce 1993)
\begin{equation}
T_{\rm dust} = 278 \left( \frac{L_\ast}{{\rm L}_\odot} \right)^{0.2} 
                   \left( \frac{r}{\rm AU} \right)^{-0.5}~~~{\rm K}~.
\label{eq:rdust}
\end{equation}
Moreover, we characterize the infrared excess in terms of the
fractional dust luminosity $L_{\rm IR}/L_\ast$, which is a
measure of the optical thickness of the dust around the star.

Table~\ref{tab:ir} summarizes the results obtained from the
infrared excess of the stars whereas Fig.\,\ref{ex_plot}
shows it graphically. The stellar parameters are taken from
Paunzen et al. (2002) and the stellar radius is calculated from
the effective temperature and luminosity of each star. 
In the case of HD\,110411, IRAS gave systematically lower fluxes
compared to ISO. Hence, we fitted the data separately. Fajardo-Acosta et al. (1999)
used a combination of the IRAS 12 and 25~$\mu$m fluxes and the
ISO 11.5 and 20~$\mu$m fluxes and fitted the infrared excess 
using a 190~K blackbody. They obtained a fractional dust luminosity 
of $5 \times 10^{-5}$. We have now an additional flux determination 
at 60~$\mu$m and obtain a dust temperature of 145~K and a
fractional dust luminosity of $1.7 \times 10^{-4}$ from the ISO data.

Vega-type stars have typical fractional dust luminosities
in the range $10^{-6}$ to $10^{-3}$ and characteristic dust
temperatures below 100~K (Backman \& Gillett 1987). Vega itself has a 
fractional luminosity of $1.5 \times 10^{-5}$ and a dust
temperature of 74~K, while $\beta$~Pictoris has $L_{\rm IR}/L_\ast = 
2.4 \times 10^{-3}$ and $T_{\rm dust}=106$~K. The fractional dust 
luminosities derived for the four $\lambda$~Bootis stars
are of the same order as those of typical Vega-type stars,
but the dust temperatures are often in excess of 100~K.

The dust around these four $\lambda$~Bootis stars is typically
located in the planet forming regions, 6 to 50~AU, but the
nature of this dust remains unknown. From the present observations,
we cannot conclude whether the dust is confined to a disk or
distributed spherically around the stars. Since we lack
detailed infrared spectroscopy at long wavelength 
($\lambda > 10 \mu$m), nothing is known about the composition
and size of these dust grains.

In a different paper, Kamp \& Paunzen (2002) speculated that normal
stars are turned for some time into $\lambda$~Bootis stars
by their passage through a diffuse interstellar cloud. In order
to elaborate on this, we calculate the radiative equilibrium radii
of typical ISM grains from Backman \& Paresce (1993)
\begin{equation}
T_{\rm dust} = 636 \left( \frac{L_\ast}{{\rm L}_\odot} \right)^{\frac{2}{11}} 
                   \left( \frac{r}{\rm AU} \right)^{-\frac{4}{11}}
                   \left( \frac{T_\ast}{{\rm T}_\odot} \right)^{\frac{3}{11}}~~~{\rm K}~,
\end{equation}
where T$_\odot$ is 5770~K and present them also in Table~\ref{tab:ir}.
The resulting radii are of the same order as the avoidance radii
of dust grains, that is the radius of closest approach for ISM
dust grains subject to the stellar radiation pressure. In this
case the infrared excess would be caused by the dust grains of
the diffuse ISM cloud as they approach the star and are repelled
by its radiation pressure.

\section{Conclusions}

From
the literature (King 1994, Fajardo-Acosta et al. 1999) it is already
clear that 1) only very few \LB members were detected in the IR and
2) even less objects show infrared excesses. 

We present in this paper all available data for members of the \LB group in the wavelength
region beyond 7000\,\AA. The data include spectrophotometry, Johnson $RIJHKLM$ and
13-color photometry as well as IRAS and ISO measurements. All the data are homogeneously
reduced and transformed into a standard system given in units of Jansky. In total,
measurements for 34 (26 with data redward of 20\,$\mu$m) well established \LB stars 
are available. From those 26 objects, 6 show an infrared excess and 2 doubtful
detections resulting in a percentage of 23\,\%.

For four stars the infrared excess was detected at two different wavelengths allows to
fit the data by a simple model. The derived dust temperatures lie between 70 and 200\,K 
and the fractional dust luminosities range from $2.2 \times 10^{-5}$ to $4.3 \times 10^{-4}$.
These values are comparable with those found for Vega-type objects. While large
black body grains indicate an origin of the IS emission in the planet forming region,
small IS dust grains point towards much larger distances (a few hundred to a few
thousand AU) from the star.

ISO-SWS spectroscopy for HD~125162 and HD~192640 resulted in the detection
of pure stellar H\,{\sc i} lines ruling out an active accretion disk around these
objects. 

Furthermore, a search for the CO (2\,$-$\,1) line at 238.538\,GHz and continuum observations
at 347\,GHz of three \LB stars yield only upper limits.

Our results clearly show that at least 23\,\% of the \LB stars have circumstellar 
or interstellar material around them. From the data summarized in this paper 
we infer that the
presence of gas and dust around the stars is not correlated. However, to draw
firm conclusions, more sensitive IR 
measurements are needed to be able to detect the stellar photosphere and thus
exclude even the presence of weak dust emission. In addition, we would need a dedicated high
resolution, high signal to noise survey for CS lines in \LB stars to detect gas in
the stellar surrounding.

\begin{acknowledgements}
We appreciate the assistance of the HHT staff and especially thank Maria Messineo
for her help with the HHT observations. We would like to thank our referee,
Dr. Turcotte for helpful comments.
This work benefitted from the Fonds zur F\"orderung der
wissenschaftlichen Forschung, project P14984.
Use was made of the SIMBAD database, operated at CDS, Strasbourg, France and
the GCPD database, operated at the Institute of Astronomy of the University
of Lausanne. The ISOPHOT data presented in this paper were reduced using PIA, 
which is a joint development by the ESA Astrophysics Division and the ISOPHOT 
Consortium with the collaboration of the Infrared Processing and Analysis 
Center (IPAC). Contributing ISOPHOT Consortium institutes are DIAS, RAL, AIP, MPIK, 
and MPIA. The ISO Spectral Analysis Package (ISAP) is a joint development by the LWS 
and SWS Instrument Teams and Data Centers. Contributing institutes are CESR, IAS, 
IPAC, MPE, RAL and SRON.
\end{acknowledgements}

\end{document}